\documentclass[12pt]{article}

\usepackage[dvips]{graphicx}

\usepackage{psfig}
\usepackage{amssymb}

\newcommand{\bF}{\mathbf{F}}
\newcommand{\bu}{\mathbf{u}}
\newcommand{\bb}{\mathbf{b}}
\newcommand{\be}{\mathbf{e}}
\newcommand{\cA}{{\mathcal A}}

\newcommand{\hu}{\hat{u}}
\newcommand{\hv}{\hat{v}}
\newcommand{\hw}{\hat{w}}
\newcommand{\hf}{\hat{f}}
\newcommand{\hg}{\hat{g}}
\newcommand{\hh}{\hat{h}}

\newcommand{\vu}{\vec{u}}
\newcommand{\vf}{\vec{f}}
\newcommand{\vb}{\vec{b}}
\newcommand{\vr}{\vec{r}}

\newcommand{\RR}{\mathbb{R}}

\newcommand{\rf}[1]{(\ref{#1})}
\newcommand{\diag}{\mbox{diag}}

\newcommand{\bivec}[2]{\left(\begin{array}{c}{#1} \\ {#2}\end{array}\right)}
\newcommand{\trivec}[3]{\left(\begin{array}{c}{#1} \\ {#2} \\ {#3}\end{array}\right)}
\newcommand{\xtext}[1]{\mbox{#1}}

\title{Computation of Strained Epitaxial Growth in Three Dimensions by
Kinetic Monte Carlo}

\author{Giovanni Russo\thanks{Dipartimento di
Matematica e Informatica, Universit\`a di Catania, Catania, Italy} \and
Peter Smereka\thanks{Department of Mathematics,
University of Michigan and the Michigan Center for Theoretical Physics,
Ann Arbor, MI 48109-1109}}

\begin{document}

\maketitle
\abstract{
A numerical method for computation of heteroepitaxial
growth in the presence of strain is presented.
The model used is based on a solid-on-solid model with a cubic
lattice. Elastic effects are incorporated using a ball and spring
type model. The growing film is evolved using Kinetic Monte Carlo (KMC)
and it is assumed that the film is in mechanical equilibrium.
The strain field in the substrate is computed by an exact
solution which is efficiently evaluated using the fast Fourier transform.
The strain field in the growing film is computed directly. The resulting
coupled system is solved iteratively using the conjugate gradient method.
Finally we introduce various approximations in the implementation of KMC
to improve the computation speed. Numerical results show that layer-by-layer
growth is unstable if the misfit is large enough resulting in the formation
of three dimensional islands.
}

\section{Introduction}

Epitaxial growth is the process where crystals are grown by the
deposition of atoms in a vacuum. Typically the deposition rate is
small and the crystal is grown, loosely speaking, one layer at a time.
In this paper, we consider the computation of strained epitaxial
growth when the strain arises because the natural lattice spacing of
the substrate and the deposited material are different. This
difference is called the {\em mismatch}. Heteroepitaxial growth is
experimentally observed to grow in the following growth modes:
\begin{enumerate}
\item
Frank-Van der Meer growth: crystal surface remains fairly flat,
growth occurs in the layer-by-layer fashion.
\item
Volmer-Weber growth:  three dimensional islands form on the
substrate without a wetting layer.
\item
Stranski-Krastanov growth: the film  grows in a layer-by- layer fashion
for a few layers, and then Volmer-Weber growth begins. This
results in three dimensional islands on top of a wetting layer.
\end{enumerate}

The type of growth mode depends on many parameters,  an important
one is the mismatch. In many cases, when the mismatch is high, one finds
Volmer-Weber growth and when the mismatch is small layer-by-layer growth
is observed. For intermediate values of the mismatch Stranski-Krastanov growth
is often seen. For an overview see, for example, Ref. \cite{gen1,gen2}.

In homoepitaxy, the effects of strain are usually very small
and quite often ignored in the many models. In general, the morphology of
a growing film  by homoepitaxy is reasonably well understood.
It is known that in some cases a homoepitaxially grown film
can undergo an instability resulting in mound formation. Typically these
phenomena are due
to kinetic effects, for example a step-edge barrier \cite{amar,john,KPM} or enhanced
edge diffusion \cite{RWG} can cause mound formation.

\begin{figure}
\label{fg:strained}
  \begin{picture}(100,220)
  \put(0,100){\centerline{\psfig{figure=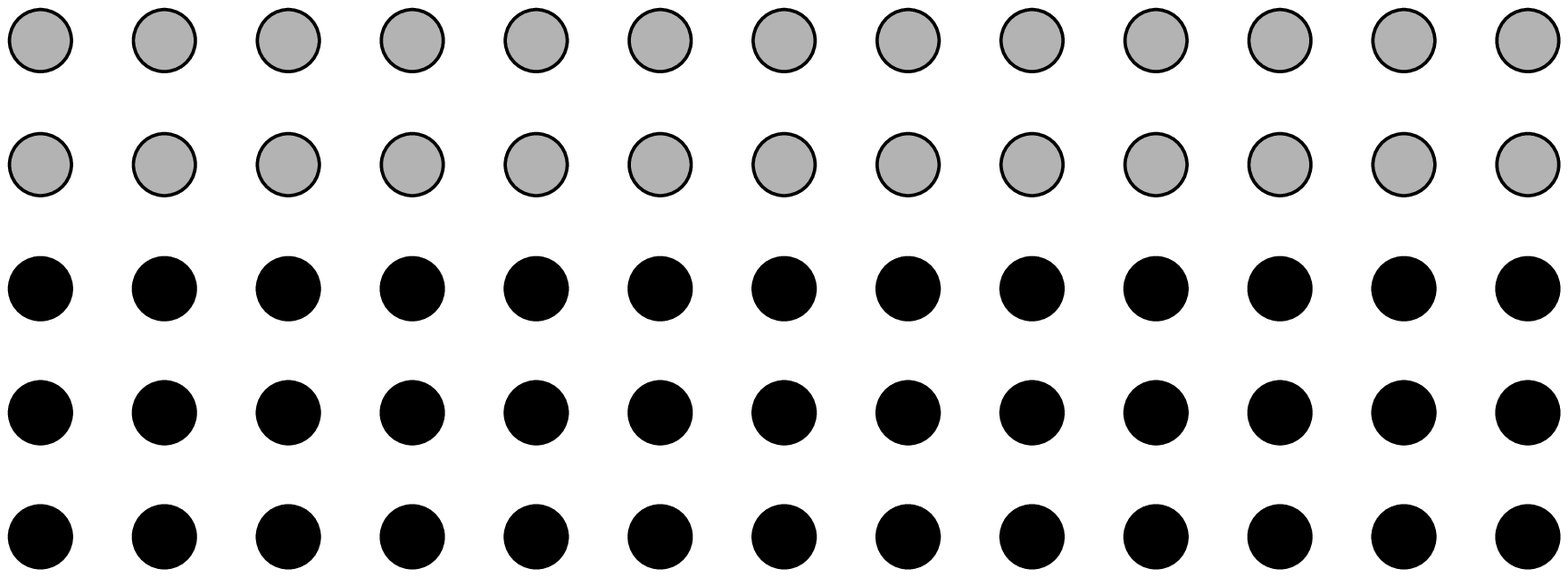,width=2.0in}}}
  \put(0,0){\centerline{\psfig{figure=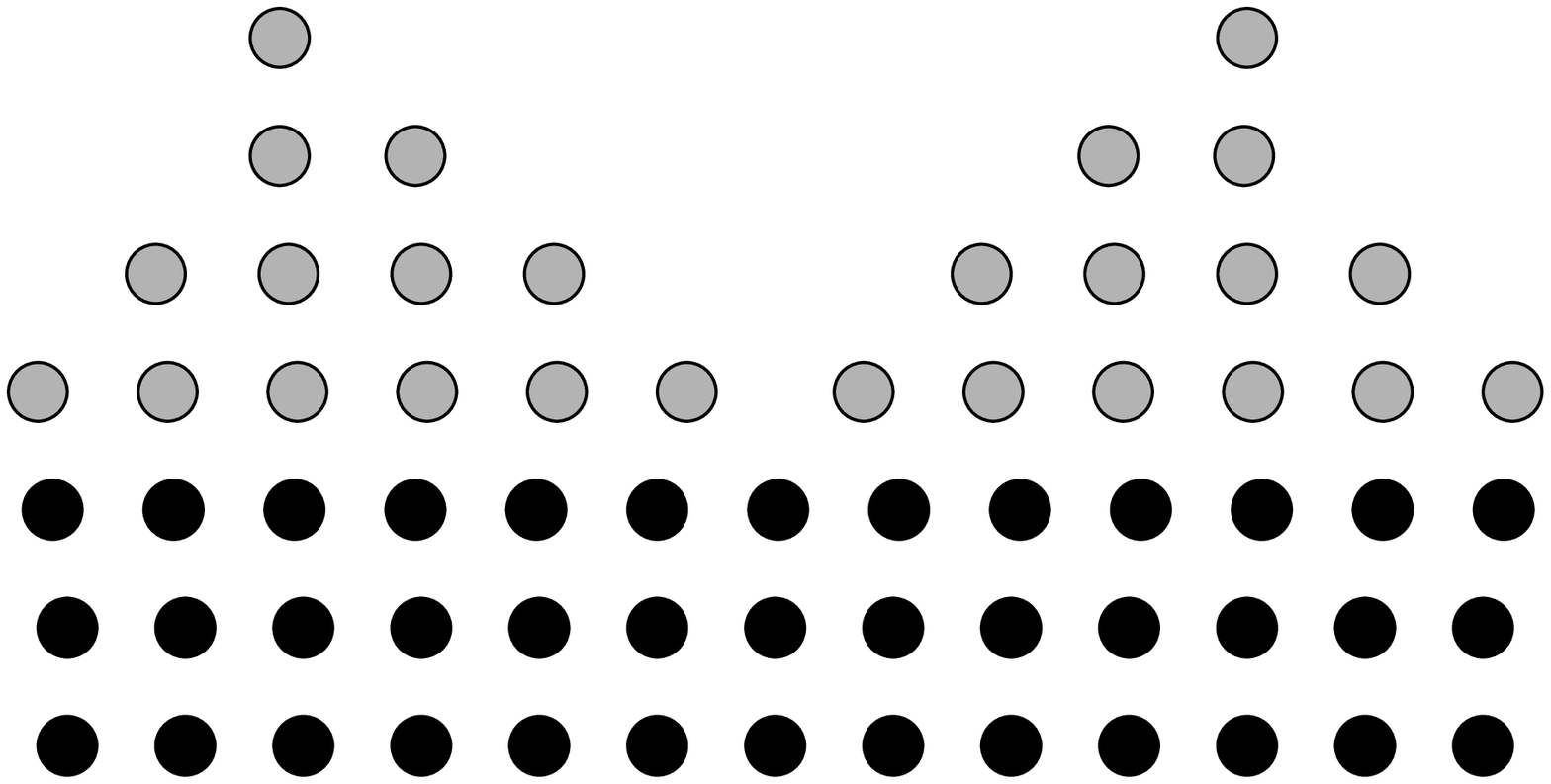,width=2.0in}}}
\end{picture}
\caption{
Germanium on Silicon-- Due to elastic interaction.
The bottom configuration has less energy than the top one}
\end{figure}

However, when a species of atoms grows on a substrate of a different
species, formation of 3D islands is observed in many situations. It is generally believed
in many cases (for example for the growth of Germanium on
Silicon) that this is a thermodynamical effect.
In particular, the  elastic energy stored in a strained flat interface
is greater than when there are three dimensional
islands. This is due to the fact that in the latter case the atoms
have more opportunity to relax (see Figure \ref{fg:strained}).
However, the surface energy
of three dimensional islands is greater than that of a flat interface. This implies
that the morphology of heteroepitaxially grown films is determined by the
interplay between elastic energy which is a bulk effect
and surface energy which arises from broken bonds.

\subsection{Modeling Elastic Effects}

Elastic effects in thin films can be studied with fully continuum
models or Burton-Cabrera-Frank \cite{BCF} type models that consider elastic effects
between steps. In this paper, we shall consider a fully discrete
model which is evolved in time using a kinetic Monte Carlo method.
Naturally, such an approach has the disadvantage of not being
able to simulate on large length scales. However, it offers
the advantage that nanoscale physical effects
such as island shape fluctuations and nucleation are naturally
incorporated. One of the first, if not the first, model in this
direction is due to Orr et al. \cite{OKSS}. They accounted for the elastic interactions
using a ball and spring model, which takes into account nearest
neighbor and next nearest neighbor interactions. This was combined with
a solid-on-solid type model which was then used with KMC to simulate
a growing heteroepitaxial film in 1+1 dimensions. If the misfit was below
a critical value, the film grew in a layer-by-layer
fashion. On the other hand, if
the misfit was above the same critical value, then the film was
observed to grow in the Volmer-Weber fashion.
Later Lam, Lee and Sander \cite{LLS} provided a more efficient
implementation of this model, which allowed them to perform
simulations using parameter values that were more physically
reasonable and compute for larger domains. This work has been recently
been extended to three dimensions \cite{LLS2}.

Ratsch et al.\ \cite{RSVZ} studied three dimensional heteroepitaxy,
however they did not take explicitly into account the harmonic forces
between atoms, but rather they used an approximate treatment
\cite{RZ1} based on the Frenkel-Kontorova model.
The model was used to investigate the island size distribution in heteroepitaxial
growth \cite{RZS}.

Off lattice KMC simulations of heteroepitaxial growth in 1+1
dimensions where presented in a series of papers
\cite{off1,off2,off3}.  In these computations the forces between atoms
were modelled using Lennard-Jones interactions. The misfit is easily
incorporated by changing parameters in the potential. One advantage of
this approach is that dislocations are naturally included, which is
not the case with the ball and spring model. These simulations also
demonstrate that if the misfit is sufficiently large, layer-by-layer
growth is unstable and mounds form.

A more sophisticated discrete elastic model was introduced by
Schindler et al. \cite{Caflisch1}. This model is based on a discrete
form of the continuum elasticity equations. The approach presented
here could be used to solve their model as well.

\section{Model description and Kinetic Monte Carlo}
\label{sec:model}

The model we shall use is a three dimensional version of the model
proposed in Refs. \cite{OKSS,LLS}. For the convenience of the reader
we shall now describe this model. To fix ideas we shall assume that
the deposited atoms are Germanium and the substrate is composed of
Silicon. The atoms occupy sites arranged on a simple cubic lattice
with no over hanging atoms allowed. This means that the height of the
surface is a function of the substrate location. We assume that atoms
bond with their nearest and next nearest neighbors.  Each atom can be
linked to its six nearest neighbors located at a distance $a$, and to
its twelve next neighbors located at a distance $a\sqrt{2}$. For
example, an atom on a flat plane surface orthogonal to one of the
coordinate axis will have five bonds with nearest neighbors, and eight
bonds with next nearest neighbors, while an atom sitting on top of
that same flat surface will have five bonds (one with a nearest
neighbor and four with next near neighbors).  We shall assume the
chemical energy associated to all these bonds is the same.  The total
chemical bond energy associated to each atom is therefore $E_b =
-\gamma N_b$, where $N_b$ is the number of bonds of each atom,
and $\gamma$ the energy associated to each bond.

The elastic effects in this model are taken into account by assuming that
the bonds will act like a spring between the atoms.
We will use $a_s$ and $a_g$ to denote the lattice spacing between
Silicon and Germanium atoms respectively.
We shall denote
respectively by $k_L$ and $k_D$ the spring constants corresponding to
longitudinal (nearest neighbor) and diagonal (next nearest neighbor) bonds.
For ease of exposition, we shall assume that both Silicon and Germanium
have the same spring constants. Since $a_g\ne a_s$ mechanical, force
will arises (the calculation of which is described in detail below).
In many systems, the time taken for sound waves to propagate across
the sample is much smaller than the time scales associated to
any growth process. Therefore we  assume that our mass-spring
system is always in mechanical equilibrium.

Each atom, $p$, of the system will hop with a rate $\Gamma_m$ \cite{LLS}
given by
\begin{equation}
   \Gamma_m = R_0\exp\left(\frac{-\Delta E}{k_BT}\right)\label{RATE}
\end{equation}
where
\begin{equation}
\Delta E = E({\rm without\ atom\ } p )-E({\rm with\ atom\ } p)\label{EEE}
\end{equation}
is the change in energy of the entire system when
atom $p$ is completely removed.  $R_0$ is the attempt frequency,
$k_B$ is the Boltzmann factor, and $T$ is the lattice temperature. Since
the chemical bonds are purely local, then we can write (\ref{EEE}) as
\[
\Delta E = n_b\gamma-\Delta E_{elas}
\]
where
$n_b$ is the number of chemical bonds of the atom,
$\gamma$ is the energy associated to the chemical bond, and
\begin{equation}
\Delta E_{elas}
= E_{elas}({\rm with\ atom\ } p)-E_{elas}({\rm without\ atom\ } p)\label{DELAS}
\end{equation}
We note that $\Delta E_{elas}$ is always nonnegative and when combined
with ({\ref{EEE}) implies that elastic effects will always increase the hopping rate.

We shall evolve the model in time by the use of
kinetic Monte Carlo (KMC).  The basic KMC method can be
described as follows.

\begin{enumerate}
\item{Pick an atom at random on the surface, uniformly among all surface atoms}
\item{Compute its number of bonds}
\item{Compute the contribution of the elastic energy associated to the atom}
\item{Pick a random number, $r$, uniformly distributed in $[0,1]$.}
\item{If $r<\exp(-(n_b \gamma - \Delta E_{elas} -K)/k_BT)$
then move the atom, according to a random
direction, chosen uniformly among all possible directions}
\end{enumerate}

The constant $K$ is computed
in such a way that the numerator $n_b \gamma - \Delta E_{elas}- K$ that
appears in the exponential is always non negative. This sets the
fastest rate in the problem to unity. The method described is
based on rejection. It is well known that if the number of possible
event types is small (as is the case for this model when there are no
elastic effects) then rejection-free Kinetic Monte Carlo
can provide a much more efficient algorithm.

While this model is idealized, it nevertheless captures the
essential physicals effects of heteroepitaxial growth, such as
adatom diffusion, nucleation, surface diffusion, long range elastic
interaction. In addition, since the model is evolved in time
using kinetic Monte Carlo, it naturally captures effects associated
with fluctuations.

\section{Elastic computations}

The main difficulty in the implementation of this model is the
computation of the strain field. In this section we shall outline
our approach for solving this problem. For the basic set up
we follow Lam et al. \cite{LLS}, as is described in next section.
However, our numerical implementation is different from the
one used in \cite{LLS}. One important feature of our work is
that we provide an exact solution for the elastic displacement in the
substrate  which is efficiently evaluated using fast Fourier transforms.
\begin{figure}
\begin{center}

\begin{picture}(0,120)
\put(-90,000){{\psfig{figure=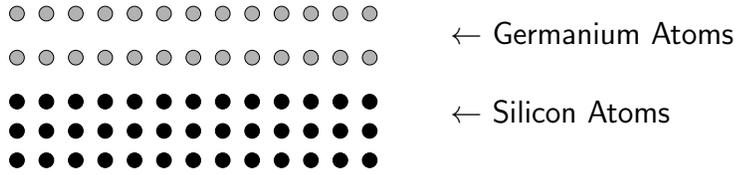,width=2.0in}}}
\put(80,60){$\leftarrow$ {\sf Germanium Atoms}}
\put(80,30){$\leftarrow$ {\sf Silicon Atoms}}
\end{picture}

\caption{The reference configuration is obtained by compressing the
Germanium atoms to have the same horizontal spacing as the Silicon
atoms.  The vertical spacing is chosen so that
Germanium atoms are in equilibrium.}
\end{center}\label{fg:refconf}
\end{figure}

\begin{figure}
\begin{center}
\includegraphics[scale=0.5]{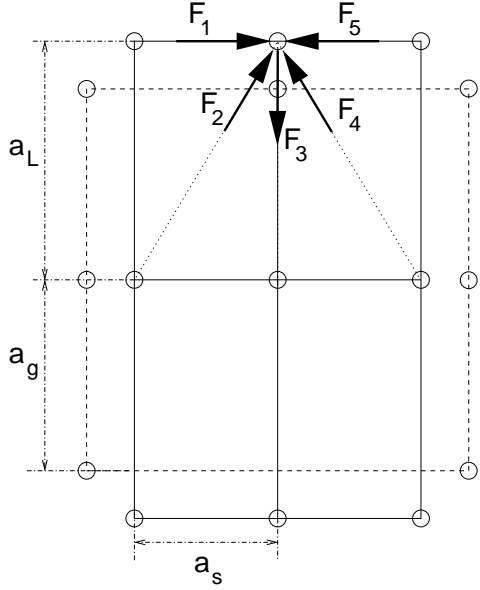}
\caption{The reference configuration is obtained by compressing the
Germanium atoms to have the same horizontal spacing as the Silicon
atoms.  The vertical spacing is chosen so that complete layers of
Germanium are in equilibrium. $\vec{b}$ is the net force on an atom due
to the compression.
In principle $\vec{b}$ can be nonzero for any
Germanium or Silicon atoms in the top row, if the top layer is not complete}
\end{center}
\label{fg:stressed}
\end{figure}

\subsection{The reference configuration}

The reference configuration we choose consists of a periodic array of
complete layers of Germanium atoms on top of a periodic array of
Silicon. The Germanium atoms are compressed so that their horizontal
lattice spacing matches that of the Silicon atoms, see Figure
\ref{fg:stressed}.
The vertical
lattice spacing, $a_L$, is chosen so that the resulting system is in mechanical
equilibrium. We will now describe the computation of $a_L$ in two dimensions. It is
useful to introduce the following dimensionless quantity
\[
   \epsilon = \frac{a_g-a_s}{a_g}.
\]
which is denoted as the mismatch. Typical values of $\epsilon$ range
from -0.05 to 0.05. For example the mismatch for Germanium
on a Silicon substrate is 0.04.
In order to deduce the atom displacement with respect to the reference
configuration we need to compute the forces experienced by an
atom due to each of its neighbors.
Elementary considerations allow to compute that, to first order in the
ratio $\epsilon$, one has
\[
    \vec{F}_1 = F_H
              \left(\begin{array}{cc} 1\\0\end{array}\right), \
    \vec{F}_2 = F_{DV}
              \left(\begin{array}{cc} 1 \\ 1\end{array}\right), \
    \vec{F}_3 = F_V
              \left(\begin{array}{cc} 0\\ -1\end{array}\right), \
    \vec{F}_4 = F_{DV}
              \left(\begin{array}{cc} -1 \\ 1 \end{array}\right), \
     \vec{F}_5=-\vec{F}_1,
\]
where $F_{V}=k_L(a_L-a_s)$, $F_H=k_L(a_g-a_s)$, and $F_{DV}=k_D(2a_g-a_L-a_s)/2$.

The value of $a_L$ is determined by requiring that these five forces
sum to zero for atoms in the reference configuration. By symmetry,
the forces in the $x$ direction sum to zero. On the other hand
balancing the $z$  components of the force one has $2F_{DV}=F_V$
which implies
\[
   k_D(2 a_g-a_L-a_s) + k_L(a_g-a_L) = 0,
\]
which gives the following expression for $a_L$
\[
   a_L=a_g \left(1 + \epsilon \frac{k_D}{k_L+k_D}\right).
\]

A similar argument can be applied to the three dimensional lattice. In
this case each atom can interact with six nearest
neighbors, located at a distance $a$, and twelve diagonal next to nearest neighbors,
located at distance $a\sqrt{2}$. The interaction with the $8$ corner
neighbors, located at a distance $a\sqrt{3}$, are neglected.

As in the two dimensional case, we shall denote by $k_L$ and $k_D$ the
two spring constants corresponding to the interaction between nearest
neighbors and diagonal neighbors.
Each bulk atom is surrounded by 18 neighbors, (six longitudinal and twelve diagonal).
We denote by  $\vec{F}_{ijk}$ the contribution of the force on a given atom due to the presence
of its neighbor in the direction $(i,j,k)$. For example, the 3D equivalent of force $\vec{F}_3$
of Figure  \ref{fg:refconf} would be $\vec{F}_{0,0,-1}$.

The six forces aligned along the coordinate axis have the expression
\begin{equation}
   \vec{F}_{ijk} = \trivec{-iF_H}{-jF_H}{kF_V},
\quad\xtext{with}\quad
i,j,k\in\{-1,0,1\},\quad\xtext{where}\quad |i|+|j|+|k|=1.\label{vertf}
\end{equation}
$F_V$ and $F_H$ are given above. The twelve diagonal forces are given by
\begin{equation}
   \vec{F}_{i0k} =-\trivec{iF_{DV}}{0}{kF_{DV}},
\quad\xtext{with}\quad
   i,k\in\{-1,1\},\label{DF1}
\end{equation}
\begin{equation}
   \vec{F}_{0jk} =-\trivec{0}{jF_{DV}}{kF_{DV}},
\quad\xtext{with}\quad
   i,k\in\{-1,1\},\label{DF2}
\end{equation}
and
\begin{equation}
   \vec{F}_{ij0} =-\trivec{iF_{DH}}{jF_{DH}}{0},
\quad\xtext{with}\quad
   i,j\in\{-1,1\}, \label{DF3}
\end{equation}
$F_{DV}$ is given above and $F_{DH}= k_D(a_g-a_s)$.
It is convenient to set $ \vec{F}_{ijk} =0$ if
$|i|+|j|+|k|=3$.

As in the two dimensional case these forces must sum to
zero in the reference configuration. The $x$ and $y$ components
will vanish by symmetry. The forces in the $z$ direction vanish
if
\[
F_V+4F_{DV}=0,
\]
which implies
\[
   2 k_D(2 a_g-a_L-a_s) + k_L(a_g-a_L) = 0.
\]
This gives the following expression for $a_L$,
\begin{equation}
   a_L= a_g\left(1+ \epsilon \frac{2k_D}{k_L+2k_D}\right).\label{AL}
\end{equation}

\subsection{Computation of the interaction}

Let us denote the
displacement, with respect to the reference configuration,
of an atom at site $(\ell, j,k)$ by the vector
$(u_{\ell jk},v_{\ell jk},w_{\ell jk})$ and the force experienced by this
atom as $(f_{\ell jk},g_{\ell jk},h_{\ell jk})$. This force will arise from the
interaction of the atom with its nearest neighbors and next nearest neighbors.
For example the $x$ component of the force is given by

\begin{eqnarray}
f_{\ell jk} &=& k_L\left(
               [u_{\ell+1jk}-u_{\ell jk}]+[u_{\ell-1jk}-u_{\ell jk}]\right)\nonumber\\
           &+&\frac{k_D}{2}\left(
               [u_{\ell+1jk+1}-u_{\ell jk}]+[u_{\ell-1jk+1}-u_{\ell jk}]\right)\nonumber\\
           &+&\frac{k_D}{2}\left(
               [u_{\ell+1jk-1}-u_{\ell jk}]+[u_{\ell-1jk-1}-u_{\ell jk}]\right)\nonumber\\
           &+&\frac{k_D}{2}\left(
               [u_{\ell+1j+1k}-u_{\ell jk}]+[u_{\ell-1j+1k}-u_{\ell jk}]\right)\nonumber\\
           &+&\frac{k_D}{2}\left(
               [u_{\ell+1j-1k}-u_{\ell jk}]+[u_{\ell-1j-1k}-u_{\ell jk}]\right)\nonumber\\
           &+&\frac{k_D}{2}\left(
               [v_{\ell+1j+1k}-v_{\ell jk}]+[v_{\ell-1j-1k}-v_{\ell jk}]\right)\nonumber\\
           &-&\frac{k_D}{2}\left(
               [v_{\ell+1j-1k}-v_{\ell jk}]+[v_{\ell-1j+1k}-v_{\ell jk}]\right)\nonumber\\
           &+&\frac{k_D}{2}\left(
               [w_{\ell+1jk+1}-w_{\ell jk}]+[w_{\ell-1jk-1}-w_{\ell jk}]\right)\nonumber\\
           &-&\frac{k_D}{2}\left(
               [w_{\ell+1jk-1}-w_{\ell jk}]+[w_{\ell-1jk+1}-u_{\ell jk}]\right)\nonumber\\
           &+&
\sum_{(m,n,q)\in{\rm neigh}(\ell,j,k)} \vec{F}_{m n q}\cdot \be_x\label{FX}
\end{eqnarray}
where $\vec{F}_{m n q}$ are given by (\ref{vertf}-\ref{DF3}). Each term in square brackets
and each $\vec{F}_{m n q}$ represents the interaction of an atom at site
$(\ell, j,k)$ with potential nearest and next nearest neighbors. If no such neighbor
exists then the term should not be included.

Suppose we have $N$ atoms and we denote the relative displacement of the $p$th atom
by $\vec{u}_p$ and let $\vec{F}_p$ denote the force it experiences. We also
let $\vb_p$ denote the sum of all forces given by
(\ref{vertf}-\ref{DF3}), acting on the atom when its position is
the reference configuration. Next we define the following vectors
in $\RR^{3N}$, $\bu=(\vec{u}_1,\ldots,\vec{u}_N)^T$, $\bb$ and $\bF$
are similarly defined. Then we can write
\[
\bF = \cA \bu + \bb.
\]
We remark that for atoms that are completely surrounded by other 18
atoms, or atoms that are on a horizontal surface, the corresponding
$\vb$ is zero, since all the forces acting on them sum up to zero, which is consistent
to the fact that a rectangular box of atoms in the
reference configuration is in equilibrium.
As a consequence, the vector $\bb$ has nonzero elements only for atoms at the surface.
The matrix vector product, $\cA\bu$ can be deduced from (\ref{FX}) and similar relations
for $g_{\ell jk}$ and $h_{\ell jk}$.

The equilibrium position of atoms in a given
configuration is obtained by setting $\bF=0$, i.e. by solving the
large linear system
\[
  \cA \bu + \bb = 0.
\]

\subsection{Contribution of the substrate}

It is known that elastic interactions can be very long
ranged. For example the elastic interaction between two
island behaves like $d^{-2}$ where $d$ is the distance between
the island centers \cite{RS}. This indicates that elastic interaction
can penetrate deep into the substrate. On the other
hand, the interaction range is certainly much shorter than
the thickness of the substrate. For this reason it is prudent to consider the
substrate to be semi-infinite in the $z$-direction. To
reduce boundary effects we consider periodic boundary
conditions in both the $x$ and $y$ directions.  In this section we shall
derive a formula that expresses the force on the surface
atoms of the substrate completely in terms of their
displacement.

The surface of the substrate corresponds to $k=0$, and
the atoms of the bulk substrate will be indexed using negative $k$
values. In the substrate $(k\leq-1)$ all atoms have a complete
set of neighbors, consequently we can write the
force, in component form, on the atom at site $(\ell, j,k)$,
$k\leq-1$, as

\begin{eqnarray}
  f_{\ell j k} & = & k_L (u_{\ell+1 j k} - 2 u_{\ell j k} + u_{\ell-1 j k}) \nonumber\\
               &   & +\frac{k_D}{2} (u_{\ell+1 j k+1} + u_{\ell-1 j k+ 1}
                     + u_{\ell+1 j k-1}+u_{\ell-1 j k-1} \nonumber\\
               &   & \quad + u_{\ell+1 j +1k} + u_{\ell-1 j+1 k}
                     + u_{\ell+1 j-1 k} + u_{\ell-1 j-1 k}-8u_{\ell j k}) \nonumber\\
               &   & + \frac{k_D}{2} (v_{\ell+1 j+1 k}+v_{\ell-1 j-1 k}
                     - v_{\ell+1 j-1 k}-v_{\ell-1 j+1 k}) \nonumber \\
               &   & + \frac{k_D}{2}(w_{\ell+1 j k+1} + w_{\ell-1 j k-1}
                     - w_{\ell+1 j k-1} - w_{\ell-1 j k+1}), \label{eq:fbulk3D}
\end{eqnarray}

\begin{eqnarray}
  g_{\ell j k} & = & k_L (v_{\ell j+1 k} - 2 v_{\ell j k} + v_{\ell j-1 k}) \nonumber \\
               &   & + \frac{k_D}{2} (v_{\ell j+1 k+1} + v_{\ell j-1 k+1}
                     + v_{\ell j+1 k-1} + v_{\ell j-1 k-1} \nonumber \\
               &   & \quad + v_{\ell+1 j+1 k} + v_{\ell-1 j+1 k}
                     + v_{\ell +1 j-1 k} + v_{\ell-1 j-1 k}- 8v_{\ell j k}) \nonumber \\
               &   & + \frac{k_D}{2} (u_{\ell+1 j+1 k} + u_{\ell-1 j-1 k}
                     - u_{\ell+1 j-1 k} - u_{\ell-1 j+1 k}) \nonumber \\
               &   & + \frac{k_D}{2} (w_{\ell j+1 k+1} + w_{\ell j-1 k-1}
                     - w_{\ell j+1 k-1} - w_{\ell j-1 k+1}), \label{eq:gbulk3D}
\end{eqnarray}

\begin{eqnarray}
  h_{\ell j k} & = & k_L (w_{\ell j k+1} - 2w_{\ell j k} + w_{\ell j k-1}) \nonumber \\
               &   & + \frac{k_D}{2} (w_{\ell j+1 k+1} + w_{\ell j+1 k-1}
                     + w_{\ell j-1 k+1} + w_{\ell j-1 k-1} \nonumber \\
               &   & + w_{\ell+1 j k+1} + w_{\ell-1 j k+1}
                     + w_{\ell+1 j k-1} + w_{\ell-1 j k-1} - 8 w_{\ell j k})\nonumber \\
               &   & +\frac{k_D}{2} (u_{\ell+1 j k+1} + u_{\ell-1 j k-1}
                     - u_{\ell+1 j k-1} - u_{\ell-1 j k+1}) \nonumber \\
               &   & +\frac{k_D}{2} (v_{\ell j+1 k+1} + v_{\ell j-1 k-1}
                     - v_{\ell j+1 k-1} - v_{\ell j-1 k+1}).  \label{eq:hbulk3D}
\end{eqnarray}
At the surface of the substrate ($k=0$), one has a slightly
different expression, because there are no atoms on top, namely:
\begin{eqnarray}
f_{\ell j 0} & = & k_L            (u_{\ell+1,j,0}   - 2u_{\ell,j,0}     +   u_{\ell-1,j,0} ) \nonumber \\
             &   & + \frac{k_D}{2}(u_{\ell+1,j,-1}  +  u_{\ell-1,j,-1}
                   +               u_{\ell+1,j+1,0} +  u_{\ell-1,j+1,0}                      \nonumber \\
             &   &               + u_{\ell+1,j-1,0} +  u_{\ell-1,j-1,0} - 6 u_{\ell,j,0} )   \nonumber \\
             &   & + \frac{k_D}{2}(v_{\ell+1,j+1,0} +  v_{\ell-1,j-1,0}
                                 - v_{\ell+1,j-1,0} -  v_{\ell-1,j+1,0})                     \nonumber \\
             &   & + \frac{k_D}{2}(w_{\ell-1,j,-1}  -  w_{\ell+1,j,-1}),                      \label{SURFFX}\\[.2in]
g_{\ell j 0} & = & k_L            (v_{\ell,j+1,0}   - 2v_{\ell,j,0}     +   v_{\ell,j-1,0})  \nonumber \\
             &   & + \frac{k_D}{2}(v_{\ell+1,j+1,0} +  v_{\ell-1,j+1,0}
                   +               v_{\ell+1,j-1,0} +  v_{\ell-1,j-1,0}                      \nonumber \\
             &   & +               v_{\ell,j+1,-1}  +  v_{\ell,j-1,-1}  - 6 v_{\ell,j,0})    \nonumber \\
             &   & + \frac{k_D}{2}(u_{\ell+1,j+1,0} +  u_{\ell-1,j-1,0}
                   -               u_{\ell+1,j-1,0} -  u_{\ell-1,j+1,0})                     \nonumber \\
             &   & + \frac{k_D}{2}(w_{\ell,j-1,-1}  -  w_{\ell,j+1,-1}),                      \label{SURFFY}\\[.2in]
h_{\ell j 0} & = & k_L            (w_{\ell,j,-1}-w_{\ell,j,0})                               \nonumber \\
             &   & + \frac{k_D}{2}(w_{\ell+1,j,-1}  +  w_{\ell-1,j,-1} +    w_{\ell,j+1,-1}
                   +               w_{\ell,j-1,-1}  - 4w_{\ell,j,0})                         \nonumber \\
             &   & + \frac{k_D}{2}(u_{\ell-1,j,-1}  -  u_{\ell+1,j,-1})                      \nonumber \\
             &   & + \frac{k_D}{2} (v_{\ell,j-1,-1} - v_{\ell,j+1,-1}).                       \label{SURFFZ}
\end{eqnarray}

Let us now consider a Fourier expansion of the displacement in the $x$
and $y$ direction. The generic Fourier mode will take the form
\begin{eqnarray}
u_{\ell jk} & = & \hu_k(\xi,\eta)e^{i(\ell\xi+j\eta)}, \nonumber \\
v_{\ell jk} & = & \hv_k(\xi,\eta)e^{i(\ell\xi+j\eta)}, \label{eq:FE} \\
w_{\ell jk} & = & \hw_k(\xi,\eta)e^{i(\ell\xi+j\eta)}. \nonumber
\end{eqnarray}
By inserting this Fourier expansion in the expression of the surface
force (\ref{SURFFX}-\ref{SURFFZ}) one obtains the relations
\begin{eqnarray}
   \hf_0 & = & 2 k_L \hu_0(\cos\xi-1)
               + k_D [\hu_{-1}\cos\xi  + \hu_0(2\cos\xi\cos\eta-3)\nonumber \\
         &\phantom{=}&
               - 2 \hat{v}_0\sin\xi\sin\eta - i\hw_{-1}\sin\xi ], \nonumber \\[.2in]
   \hg_0 & = & 2 k_L \hv_0(\cos\eta-1)
               + k_D[\hv_{-1}\cos\eta  + \hv_0(2\cos\xi\cos\eta-3)\nonumber \\
         &\phantom{=}&
               - 2\hat{u}_0\sin\xi\sin\eta - i\hw_{-1}\sin\eta ]\label{eq:F1}, \\[.2in]
   \hh_0 & = &   k_L(\hw_{-1}-\hw_0) + k_D[\hw_{-1}(\cos\xi+\cos\eta)
               - 2\hw_0]\nonumber\\
         &\phantom{=}&
-i(\hu_{-1}\sin\xi + \hv_{-1}\sin\eta) ]. \nonumber
\end{eqnarray}
In the relation above we have suppressed the dependence of all Fourier modes
on $(\xi,\eta)$.

Eq.\rf{eq:F1} gives a relation between the Fourier modes of the force
and the Fourier modes of the displacement.  However, our goal is to
express $(\hf_0,\hg_0,\hh_0)$ in terms of $(\hu_0,\hv_0,\hw_0)$. Once
this is done then the force field at the surface can be computed from
its Fourier modes by inverse discrete Fourier transform.
In order to accomplish our goal, we need to express
$\hu_{-1},\hv_{-1},\hw_{-1}$ in terms of $\hu_0,\hv_0,\hw_0$, and
substitute their expression into \rf{eq:F1}.
This can be done as follows. First, let us insert the Fourier
expansion \rf{eq:FE} into (\ref{eq:fbulk3D}--\ref{eq:hbulk3D}) obtaining:

\begin{eqnarray}
   \hf_k & = & 2 k_L \hu_k (\cos\xi-1) + k_D[(\hu_{k+1}+\hu_{k-1})\cos\xi
               + \hu_k(2\cos\eta\cos\xi-4)]  \nonumber \\
         &   & + i k_D(\hw_{k+1}-\hw_{k-1})\sin\xi - 2 k_D\hv_k\sin\xi\sin\eta\nonumber \\[.2in]
   \hg_k & = & 2 k_L\hv_k (\cos\eta-1)+k_D[(\hv_{k+1}+\hv_{k-1})\cos\eta
               + \hv_k (2\cos\eta\cos\xi-4)]  \label{EDIS} \\
         &   & + i k_D(\hw_{k+1}-\hw_{k-1})\sin\eta-2k_D\hu_k\sin\xi\sin\eta\nonumber \\[.2in]
   \hh_k & = & k_L(\hw_{k+1}-2\hw_k+\hw_{k-1}) \nonumber \\
         &   & + k_D[(\hw_{k+1}+\hw_{k-1}) (\cos\xi+\cos\eta)-4\hw_k] \nonumber \\
         &   & + i k_D[(\hu_{k+1}-\hu_{k-1})\sin\xi+(\hv_{k+1}-\hv_{k-1})\sin\eta]\nonumber
\end{eqnarray}

The discrete equations given by (\ref{EDIS})  are solved using the following
substitution
\begin{equation}
  \hu_k = \hu\alpha^k,\quad  \hv_k =  \hv\alpha^k,  \quad \hw_k =  \quad \hw\alpha^k,
  \label{eq:uvwk}
\end{equation}
where we look for solutions with $|\alpha|>1$, since we expect the
Fourier modes to decay as $k\to-\infty$.  Inserting this {\em
ansatz\/} into the expression \rf{EDIS} of the Fourier modes of the force acting
on the inner points of the substrate, one obtains:
\begin{equation}
   \left(\begin{array}{c}\hf \\ \hg \\ \hh \end{array}\right) = \Omega(\alpha)
   \left(\begin{array}{c}\hu \\ \hv \\ \hw \end{array}\right),
\end{equation}
where the entries of the matrix $\Omega$ are given by
\begin{eqnarray*}
  \omega_{11} & = & 2 k_L(\cos\xi -1)\alpha + k_D[\cos \xi(1+\alpha^2)
                    + 2 (\cos\eta\cos\xi-2)\alpha] \\
  \omega_{22} & = & 2 k_L(\cos\eta-1)\alpha + k_D[\cos\eta(1+\alpha^2)
                    + 2 (\cos\eta\cos\xi-2)\alpha] \\
  \omega_{33} & = & k_L(\alpha^2-2\alpha+1) + k_D[(\alpha^2+1)(\cos\xi+\cos\eta)-4] \\
  \omega_{12} & = & \omega_{21} =  - 2\alpha k_D\sin\xi\sin\eta \\
  \omega_{13} & = & \omega_{31} = i k_D (\alpha^2-1)\sin\xi \\
  \omega_{23} & = & \omega_{32} = i k_D (\alpha^2-1)\sin\eta \\
\end{eqnarray*}
Note that matrix $\Omega$ is symmetric, but not self-adjoint.

Since all forces in the bulk have to be zero (all such atoms are in
mechanical equilibrium), then one has
\begin{equation}
   \Omega    \left(\begin{array}{c}\hu \\ \hv \\ \hw \end{array}\right) = 0.
   \label{eq:eigenval}
\end{equation}
This homogeneous system has  nontrivial solutions only if
\begin{equation}
   P(\alpha)\equiv \det(\Omega) = 0.
   \label{eq:detOmega0}
\end{equation}
This relation results in an algebraic equation for the values of
$\alpha$. The polynomial $P(\alpha)$ is a six degree polynomial,
therefore it admits, in general, six roots.  Note that, because
of the structure of the matrix $\Omega$, matrix
$\alpha^2\Omega(1/\alpha)$ is equal to $\Omega(\alpha)$ with
$\omega_{13}=\omega_{31}$ and $\omega_{23}=\omega_{32}$ of opposite sign.
This does not change the expression of the determinant, and therefore
if $\tilde \alpha \neq 0$ is a root, then also $1/\tilde\alpha$ is a root.  This means
that the number of roots $\tilde\alpha$ such that $|\tilde\alpha|>1$
is equal to the number of (nonzero) roots $\tilde\alpha$ such that
$|\tilde\alpha|<1$.  The roots that are of interest for us are the
ones that decay as $k\to\infty$, i.e. $\tilde \alpha$ : $|\tilde
\alpha|>1$.

\subsubsection{Eigenvector computation}
In this subsection we describe a general procedure for the computation
of the eigenvalues and eigenvectors, that works also in the case of
multiple eigenvalues.

The goal is to solve the problem given by (\ref{eq:eigenval}),
which we write as
\begin{equation}
\Omega r = 0\label{eq:Ei1}
\end{equation}
where $r$ is a three-component vector (we drop the arrow on top),
and to find independent eigenvectors even if some eigenvalues coincide.
First compute the eigenvalues by solving the algebraic equation
\rf{eq:detOmega0}. Consider the three eigenvalues $\alpha_\ell$ such that
$|\alpha_\ell|>1$, $\ell=1,\ldots,3$. If they are all distinct, then
the three eigenvectors corresponding to them will be independent. If
two of them are coincident, let us say $\alpha_2=\alpha_3$, then one
has to find two independent eigenvectors corresponding to the
coincident eigenvalues.

A unified treatment of the problem is obtained by the use of the
singular value decomposition (SVD) of matrix $\Omega$. The procedure works
as follows.
First compute $\alpha_1,\alpha_2,\alpha_3$. If they are distinct, for
each of them compute $\Omega_\ell = \Omega(\alpha_\ell)$, $\ell =
1,\ldots,3$.  Perform the SVD of $\Omega_\ell$: $\Omega_\ell = U\Sigma
V^\dagger$, where $U$ and $V$ are unitary matrices (i.e. $UU^\dagger =
I$, $VV^\dagger = I$), and $\Sigma$ is a diagonal matrix containing
the singular values of $\Omega_\ell$. Taking into account that $U$ is
non singular, problem \rf{eq:Ei1} reads
\[
  \Sigma V^\dagger r = 0.
\]
Since $\Omega_\ell$ is singular, then $\Sigma = \diag(\sigma_1,\sigma_2,0)$,
therefore one has
\[
  \sigma_1\left(V^\dagger r\right)_1 = 0, \quad
  \sigma_2\left(V^\dagger r\right)_2 = 0, \quad
  \left(V^\dagger r\right)_3 = \mbox{arbitrary}
\]
Let us choose $\left(V^\dagger r\right)_3 = 1$.

Assuming $\sigma_2\neq0$, i.e. that the matrix $\Sigma_\ell$ has rank 2,
then one has
\[
  V^\dagger r = \left(\begin{array}{c}0\\ 0\\ 1\end{array}\right),
\]
therefore
\[
  r = V  \left(\begin{array}{c}0\\ 0\\ 1\end{array}\right)
\]
i.e. $r$ is the third column of $V$.

If two roots are coincident, say $\alpha_1, \alpha_2=\alpha_3$, then
first compute the eigenvector $r_1$ using the procedure above applied
to matrix $\Omega(\alpha_1)$.
For the computation of the other eigenvectors there are two
possibilities: either the rank of the matrix $\Omega_2=\Omega_3$ is 1,
i.e. $\sigma_2=0$, or the rank of the matrix is 2, i.e. $\sigma_2\neq
0$. However, the latter case never happened in all our computations,
and we conjecture it can never happen for our problem. Therefore we
assume that $\sigma_2=0$. Repeating the procedure above, one finds
that $r_2$ and $r_3$ can be computed, respectively, as the second and
third column of the matrix $V$.

\subsubsection{Surface Force Formula}

We denote by $\vr_\ell$ a solution of the system
\[
  \Omega(\alpha_\ell)\vr=0,
\]
with $\ell = 1,2,3$. Then we can decompose the vector
$(\hu_0,\hv_0,\hw_0)^T$ on the basis of the eigenvectors, i.e.
\[
 \left(\begin{array}{c}\hu_0 \\ \hv_0 \\ \hw_0 \end{array}\right)
       = c_1 \vr_1 + c_2 \vr_2 + c_3 \vr_3.
\]
Once the constants $c_1,c_2,c_3$ are computed, one can write
\[
    \left(\begin{array}{c}\hu_{-1} \\ \hv_{-1} \\ \hw_{-1} \end{array}\right) =
    \frac{c_1}{\alpha_1} \vr_1 + \frac{c_2}{\alpha_2} \vr_2 + \frac{c_3}{\alpha_3} \vr_3.
\]
The two relations allow to express $\hu_{-1}$, $\hv_{-1}$, $\hw_{-1}$
in terms of $\hu_{0}$, $\hv_{0}$, $\hw_{0}$.  Once this is done, one
can substitute this expression into \rf{eq:F1} and obtain the final
relation between $(\hf_0,\hg_0,\hh_0)$ and $(\hu_0,\hv_0,\hw_0)$.
This relation has to be computed for all Fourier modes
$(\xi,\eta)$. Periodicity implies that $\xi = 2\pi m/M$, $\eta =
2\pi n/M$, $m,n = 1, \ldots, M$.

The complete algorithm for the computation of $\vf_{\ell j 0}$ from
$\vu_{\ell j 0}$ can be summarized as follows.

\centerline{\bf Computation of $\vf_{\ell j 0}$ from $\vu_{\ell j 0}$}

\begin{enumerate}
\setcounter{enumi}{-1}
\item {\it Preprocessing.} Given $M$, for each mode $(m_1,m_2), m_1,m_2=1,\ldots,M$,
solve the eigenvalue problem \rf{eq:eigenval}, and store the eigenvalues and eigenvectors.

\item Given $\vu_{\ell j 0}$, perform the discrete Fourier transform
   in $\ell$ and $j$ and compute all Fourier modes $\hu_0,\hv_0,\hw_0$.

\item For each mode, compute $\hu_{-1}, \hv_{-1}, \hw_{-1}$ using pre-computed
   values of eigenvalues and eigenvectors.

\item Compute the Fourier modes of the force $\hf,\hg,\hh$, using Eq.\rf{eq:F1}

\item Compute the force by inverse discrete Fourier transform
\end{enumerate}
All discrete Fourier transforms can be efficiently computed by FFT
algorithms in $O(M^2\log M)$ operations. In all our calculations we used
the FFTW package developed at MIT \cite{FFTW}.

\subsection{Elastic Displacement Computation}
Let us assume that we have deposited $N$ atoms on a substrate of size
$M\times M$.  Let us use
$\bu_g\in\RR^{3N}$ to denote the relative
displacement of the Germanium atoms on the
substrate. We use $\bu_s\in\RR^{3M^2}$ to denote  the
relative displacement of the top layer of atoms on the substrate.
Then the equilibrium position of the particles can be obtained by
solving the following linear system
\begin{equation}
   \label{eq:big}
   \bF \equiv \left(\begin{array}{c}\bF_s  \\ \bF_g\end{array}\right) =
              \left(\begin{array}{cc}S & B \\ B^T & A\end{array}\right)
              \left(\begin{array}{c}\bu_s  \\ \bu_g\end{array}\right) +
              \left(\begin{array}{c}\bb_s  \\ \bb_g\end{array}\right) = 0
\end{equation}
The matrices appearing in the system have the following meaning. The
forces acting on the $M^2$ Silicon atoms on the surface of
the substrate are
\[
   \bF_s = S \bu_s + B \bu_g + \bb_s.
\]
Here $S \bu_s$ is the force on the atoms at the surface of the substrate
due to {\it all\/} the (Silicon) atoms in the substrate. This is
efficiently computed using the results from the previous
section.
$B \bu_g$ is the force on the substrate surface due to the Germanium
atoms, and $\bb_s$ is the sum of the forces given by
(\ref{vertf}-\ref{DF3}). The force acting
on the $N$ Germanium atoms on the substrate is given by
\[
  \bF_g = B^T\bu_s + A \bu_g +\bb_g,
\]
where $ A \bu_g $ are the forces that arise from the
interactions between the Germanium atoms, $B^T\bu_s$ is
the force on the Germanium atoms due to the top
layer of Silicon atoms, and $\bb_g$ is the sum of the
forces given by (\ref{vertf}-\ref{DF3}).

We observe that the matrix
\begin{equation}
   \label{MMM}
  \left(\begin{array}{cc}S & B \\ B^T & A\end{array}\right)
\end{equation}
is a symmetric  negative semi-definite  matrix; it
has $3$ zero eigenvalues, corresponding to the free
translation in the $3$ directions of the coordinate axis. The system
is clearly invariant for translation along the directions parallel to
the substrate. It is also invariant along the direction orthogonal to
the substrate, because the substrate is considered semi-infinite. This
can be understood by the following argument. For a substrate of a
finite thickness, let us say of $N_L$ layers, a unit displacement in
the direction orthogonal to the substrate will produce an elastic
force per unit atom equal to
\[
   F = \frac{1}{N_L} (k_L + 2 k_D)
\]
which vanishes as $N_L\to\infty$. Therefore no resistance is opposed
to any translation.

Notice that the matrix $B^T$ is the
transpose of matrix $B\in\RR^{3M^2\times 3N}$. $A$ is a $ 3N\times 3N$
matrix. $A$ and $B$ are sparse and the matrix vector products are
efficiently evaluated using expressions similar to (\ref{FX}).
System \rf{eq:big} can be solved by an iterative scheme for large,
sparse linear systems, making use of the symmetry and definiteness of
the coefficient matrix.
Here we shall use just a simple conjugate gradient method, leaving the
search for a more efficient method to future investigations.

\section{Evaluation of the elastic energy}
Once the strain field is determined, the elastic energy is computed as follows.
The energy associated to the bonds is given by
\[
  E_{elas} = E_{\rm Ge-Ge} + E_{\rm Ge-Si} + E_{\rm Si-Si},
\]
where $E_{\rm Si-Si}$ is the energy due to the interaction between
the  Silicon atoms. The other terms are analogously defined.
One has
\begin{equation}
   E_{\rm Si-Si} =
   \sum_{\rm Si-Si\>bonds} \frac12 k_{\rm bond} (\ell_{\rm bond})^2 .
   \label{eq:Si-Si-bonds}
\end{equation}
where $ k_{\rm bond} $ is either $k_L$ or $k_D$ depending
on whether the bond is longitudinal or diagonal.
$\ell_{\rm bond}$ is the amount the bond has been stretched
from the equilibrium configuration. This can be written
in terms of the displacement field as
\begin{equation}
   E_{\rm Si-Si} =
   -\frac12 \bu_{\rm Si}^TA_{\rm Si}\bu_{\rm Si},
   \label{eq:Si-Si-bonds2}
\end{equation}
where we denote by $A_{\rm Si}$ the (infinite dimensional) matrix that
provides the force on all Silicon atoms as a function of the position
of the Silicon atoms.

The energy due to the interaction of the Germanium atoms
can be written as
\begin{equation}
   E_{\rm Ge-Ge} + E_{\rm Ge-Si} =
   \sum_{\rm all\>Ge\>bonds}\frac12 k_{\rm bond} (\ell_{\rm bond})^2 ,
   \label{eq:Ge-bonds}
\end{equation}
where $k_{\rm bond}$ is as in (\ref{eq:Si-Si-bonds}) but here
$\ell_{\rm bond}$ represents the amount the Germanium bonds have been stretched
from their original equilibrium configuration (as opposed to the
reference configuration). This can be written
as
\begin{equation}
   E_{\rm Ge-Ge} + E_{\rm Ge-Si} =
   \frac12
\sum_{\begin{array}{ccc}
{\rm{ \scriptstyle all\> atoms\>}}\\[-.08in]{\rm{\scriptstyle bonded \> to}}
\\[-.08in] {\rm{\scriptstyle a\> Ge\> atom}}
  \end{array}}\hspace{-.2in} e_{\ell jk}
   \label{eq:tope}
\end{equation}
where $e_{\ell jk}$ is the total elastic energy stored in all the
bonds associated to the atom located at site $(\ell, j,k)$. The factor
$\frac12$ accounts for the double counting of the summation.
We can write
\[
e_{\ell jk}=e^x_{\ell jk}+e^y_{\ell jk}+e^z_{\ell jk}
\]
with
\begin{eqnarray*}
e^x_{\ell jk}
 &=& \frac{k_L}{2}\left(
      [u_{\ell+1jk}-u_{\ell jk}+d_x]^2+
      [u_{\ell -1jk}-u_{\ell jk}-d_x]^2\right)\\
 &+&\frac{k_D}{2}\left(
      [u_{\ell+1jk+1}-u_{\ell jk}+d_x]^2+
      [u_{\ell -1jk+1}-u_{\ell jk}-d_x]^2\right)\\
 &+&\frac{k_D}{2}\left(
      [u_{\ell+1jk-1}-u_{\ell jk}+d_x]^2+
      [u_{\ell -1jk-1}-u_{\ell jk}-d_x]^2\right)\\
 &+&\frac{k_D}{2}\left(
      [u_{\ell+1j+1k}-u_{\ell jk}+d_x]^2+
      [u_{\ell -1j+1k}-u_{\ell jk}-d_x]^2\right)\\
 &+&\frac{k_D}{2}\left(
      [u_{\ell+1j-1k}-u_{\ell jk}+d_x]^2+
      [u_{\ell -1j-1k}-u_{\ell jk}-d_x]^2\right)
\end{eqnarray*}
where $d_x=a_s-a_g$. In the above expression, each term in square
brackets represents the contribution to the elastic energy  by a
pair of atoms. If no such pair exists then the term is not included.
One can derive analogous expressions for $e^y_{\ell jk}$ and $e^z_{\ell jk}$
where $d_y=d_x$ and $d_z=a_L-a_g$.

To compute the total elastic energy,
the sum in Eq.\rf{eq:Ge-bonds} is computed directly, while the sum in Eq.
\rf{eq:Si-Si-bonds}, which contains infinitely many terms, can be computed
by the following argument.
First, let us distinguish among the surface and the bulk
atoms. Let us denote by
$-\bF_{\rm Si}$ the force acting on Silicon due to the presence of the Germanium.
They take the form
\[
  \bF_{\rm Si} = \bivec{\bF_{s}}{0},
\]
where the dimension of the vector $\bF_s$ is equal to
the dimension of the surface of the substrate,
while $0$ represent the force acting on  rest of all the infinite atoms of the bulk.

At equilibrium, the net force acting on all Silicon atoms is zero, therefore we may write
\[
  -\bF_{\rm Si} + A_{\rm Si}\bivec{\bu_{s}} {\bu_{\rm bulk}} = 0.
\]
Using the above relations one obtains
\[
  E_{\rm Si-Si} = -\frac12\bivec{\bu_s}{\bu_{\rm bulk}}^T
  \bivec{\bF_{s}}{0} = -\frac{1}{2}\bu_{s}^T\bF_{s}
\]

We remark here that $\bF_{s}$ is the same surface force computed
in the previous section using the discrete Fourier transform.

\section{Time stepping approximations}
The method outlined in Section \ref{sec:model} is impractically slow
for the following reasons
\begin{enumerate}
\item
For each attempted hop, a complete elastic computation has to be
performed.  Since most attempts are rejected, most of the time would
be spent performing elastic computations that are never used.

\item
Even without elastic effects, the simple rejection-based KMC described
here is very slow. A more effective technique would be to use
rejection-free KMC, in which all possible events are sampled according
to their probability \cite{BKL,rand2}. However, with the inclusion of
elastic effects, the implementation of rejection-free KMC is not so
straightforward.

\end{enumerate}

Here we shall outline various approximations of the model which lead to a much
faster code, without significantly compromising the physical fidelity.

As mentioned above, in order to know the rate at which an
atom might hop we must compute the change in elastic
energy of the entire system with and without that atom
present. We make the following approximation, we assume that
the change in the elastic energy is due to the energy
in the bonds that directly connect that atom. Therefore
we have
\begin{equation}
 \Delta E_{elas}\approx
\sum_{\begin{array}{cc}
{\rm{ \scriptstyle bonds \> to}}\\[-.08in] \scriptstyle {{\rm atom} \> p}
  \end{array}}\hspace{-.2in} \frac{k_{\rm bond}}{2}(\ell_{\rm bond})^2 = e_{\ell jk}\label{eapprox}
\end{equation}
where $(\ell, j,k)$ is the site of the $p$th atom.
The advantage of this approach is that a new equilibrium configuration has to be
computed only if the move is accepted. Another
approximation involves updating the  displacement field after
a given number $J$ of hops.
Furthermore, one can verify that the change in elastic
energy for atoms which are lightly bonded ($N_b\leq5$) is
very small and consequently we assume it to be zero.

Finally, in order to reduce the number of rejections  we separate
the lightly bonded ($N_b\leq5$) and the  more strongly bonded
($N_b > 5$) atoms in our implementation of kinetic Monte Carlo.
This is done as follows. We take $Q$ steps where we update the
lightly bonded atoms and then take one step where the
strongly bonded atoms are allowed to move. The accepted rate
for the strongly bonded atoms is increased by a factor $Q$.

The above discussion can be conveniently summarized
as the following algorithm.

\bigskip

\centerline{ALGORITHM}

\begin{enumerate}
\item{Choose a site at random among all $M\times M$ sites (only atoms on the surface are allowed to move)}
\item{If there is Germanium atom present compute $N_b$ (number of bonds)}
\item{Let it hop without any elastic computation if $N_b \le 5$ }
\item{If $N_b > 5$ ignore the atom}
\item{Repeat Steps 1 to  4 $M^2$ times}
\item{Repeat Steps 1 to  5 $Q$ times}
\item{Let the system come to mechanical equilibrium (perform an elastic solve).}
\item{Choose a site at random}
\item{If there is Germanium atom present compute $N_b$}
\item{If $N_b \le 5$ do nothing }
\item{If $N_b > 5$  then compute a random number, $r\in U[0,1]$}
\item{If $r < Q\exp\left({-\gamma (N_b-5)+\Delta E_s}\right)$ perform a hop}
\item{After $J$ hops update the displacement field}
\item{Repeat Steps 9 to 13 $M^2$ times.}
\item{We have now advanced $Q$ steps}
\end{enumerate}

\section{Numerical results}
At the present time the algorithm is to too slow to perform computations
for realistic values for the parameter values and see effects
due to elastic strain. For example, quantum dots  observed in
experiments are on the order of 20 nm. This would suggest that
we should be computing on domains on the order of $1000\times 1000$.
At the present time the largest domain for which we can simulate in a reasonable time
is $ 64 \times 64$. Since elastic phenomena are a bulk effect,
then we must increase the spring constants to be unphysically large
in order to observe significant elastic interaction. In our
simulations we chose $k_L=500$ and $k_D=250$. These values
are approximately 10 times larger than physical values.
In addition we chose $F=10^{-5}$. Since the hopping rate
for an adatom is unity then it follows that the diffusion coefficient is $D=\frac14$. Therefore
in our simulations  we have $D/F=2.5\times 10^4$. From an experimental
point of view this is small. Realistic values for $D/F$
are typically larger than $10^5$. We have chosen high deposition rate to
reduce the simulation time. We have taken $\gamma= 2$. Finally we
have used $Q=5$ and $J=8$. Numerical experiments revealed taking smaller
values for $Q$ and $J$ did not change the answer appreciably.

In Figures \ref{fg:two}, \ref{fg:four}, and \ref{fg:six}, we present computations
using the parameter values discussed above but we allow the misfit to
vary. Figure \ref{fg:two} shows the growth for $\epsilon =0.02$. Here one
observes layer-by-layer growth.  Figure \ref{fg:four} presents the results
when $\epsilon =0.04$. One observes that in the initial stages of
growth the morophology is similar to layer-by-layer growth but three
dimensional islands form  by nucleation type events and by the formation
of trenches \cite{tersoff}. The result for the case $\epsilon =0.06$ are shown in Figure
\ref{fg:six}. In this situation three dimensional islands form very
quickly by nucleation.

In both cases, $\epsilon=0.04$ and $\epsilon=0.06$ we observe Volmer-Weber growth.
Simulations were performed over a wide range of parameter values and they
always revealed a sharp transition between layer-by-layer growth
and Volmer-Weber growth; Stranski-Krastanov growth was not
observed. Our results are consistent with previous simulations in this regard.

\begin{figure}
\begin{picture}(200,400)
\put(0,250){\centerline{
  \psfig{figure=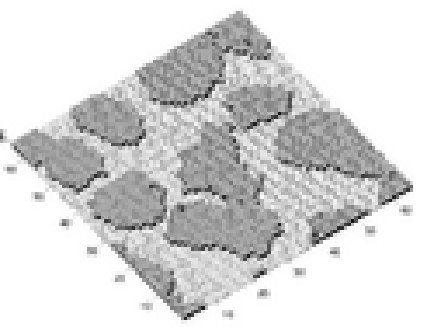,width=3in}
  \psfig{figure=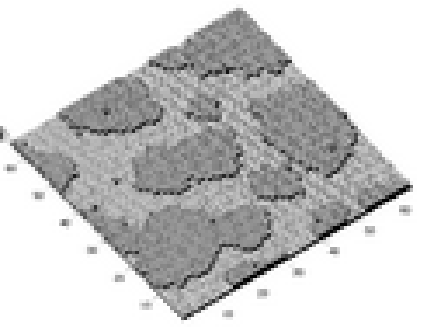,width=3in}}}

\put(0,30){\centerline{
  \psfig{figure=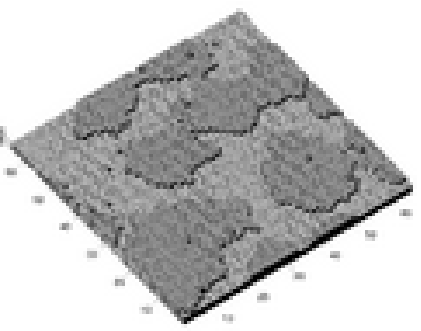,width=3in}
  \psfig{figure=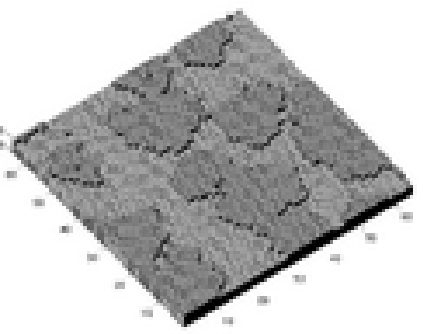,width=3in}}}

\put(120,230){(a)}
\put(350,230){(b)}
\put(120,10){(c)}
\put(350,10){(d)}

\end{picture}
  \caption{Heteroepitaxial simulations with $\epsilon = 0.02$, all other parameter values
are given in the text. (a) 0.5 monolayers, (b) 1.5 monolayers, (c) 2.5 monolayers,
and (d) 3.5 monolayers. The number of gray levels is equal to the maximum height.
Note: higer resolution versions of the figures can be found
at www.math.lsa.umich.edu/$\sim$psmereka}
\label{fg:two}
\end{figure}

\begin{figure}
\begin{picture}(200,400)
\put(0,250){\centerline{
  \psfig{figure=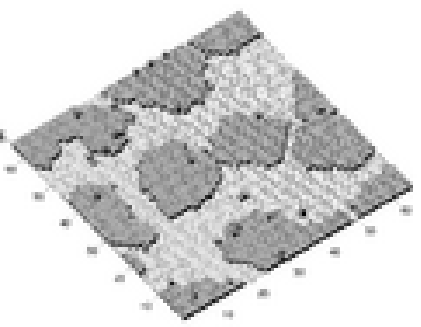,width=3in}
  \psfig{figure=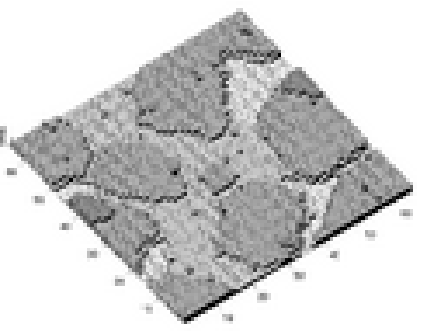,width=3in}}}

\put(0,30){\centerline{
  \psfig{figure=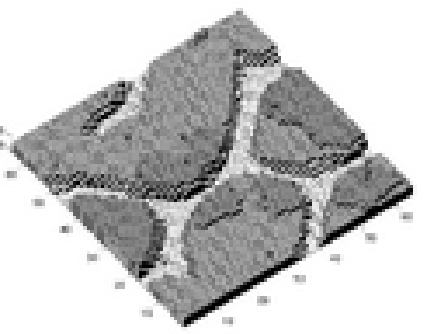,width=3in}
  \psfig{figure=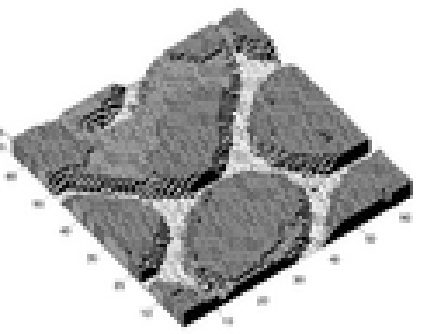,width=3in}}}

\put(120,230){(a)}
\put(350,230){(b)}
\put(120,10){(c)}
\put(350,10){(d)}

\end{picture}
  \caption{Heteroepitaxial simulations with $\epsilon = 0.04$, all other parameter values
are given in the text. (a) 0.5 monolayers, (b) 1.5 monolayers, (c) 2.5 monolayers,
and (d) 3.5 monolayers. The number of gray levels is equal to the maximum height.}
\label{fg:four}
\end{figure}

\begin{figure}
\begin{picture}(200,400)
\put(0,250){\centerline{
  \psfig{figure=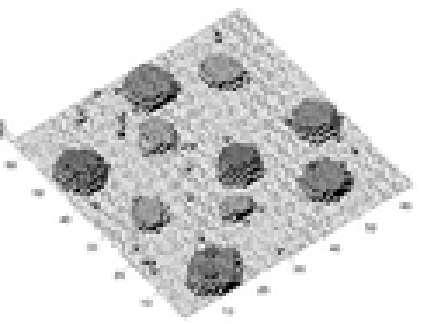,width=3in}
  \psfig{figure=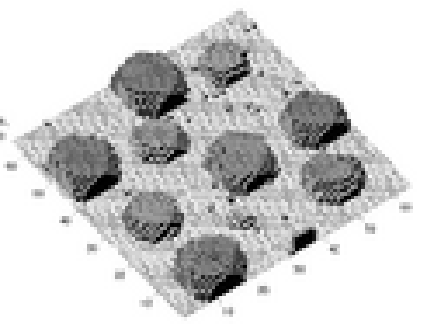,width=3in}}}

\put(0,30){\centerline{
  \psfig{figure=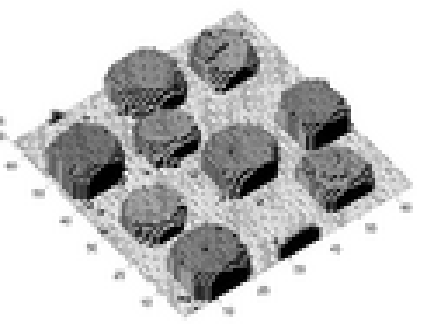,width=3in}
  \psfig{figure=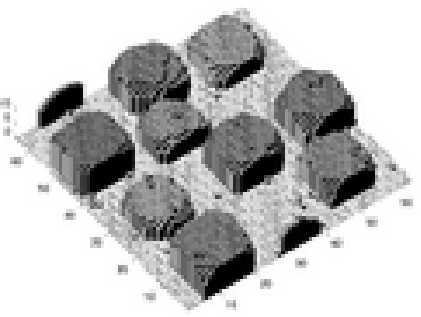,width=3in}}}

\put(120,230){(a)}
\put(350,230){(b)}
\put(120,10){(c)}
\put(350,10){(d)}

\end{picture}
\caption{Heteroepitaxial simulations with $\epsilon = 0.06$, all other parameter values
are given in the text. (a) 0.5 monolayers, (b) 1.5 monolayers, (c) 2.5 monolayers,
and (d) 3.5 monolayers. The number of gray levels is equal to the maximum height.}
\label{fg:six}
\end{figure}

\section{Summary}
A numerical method for the computation of heteroepitaxial growth in
the presence of strain using kinetic Monte Carlo has been presented.
A solid-on-solid model is used and the elastic effects are incorporated using a
linear ball and spring model. It is assumed that the film is mechanical equilibrium.
The strain field in the substrate is computed by an exact
solution which is efficiently evaluated using the fast Fourier transform.
The strain field in the growing film is computed directly. The resulting
coupled system is solved iteratively using the conjugate gradient method.
Finally we introduce various approximations in the implementation of the KMC
to improve the computation speed. Numerical results show that layer-by-layer
growth is unstable if the misfit is large enough resulting in the formation
of three dimensional islands. Our results are in agreement with
previous studies \cite{off2,off1,off3,LLS,OKSS,RSVZ}.

Currently, we are in the process of solving the elastic equations
for the deposited atoms using multigrid and then coupling
the multigrid solver to the exact solution in the substrate.
In addition we plan to extend the model to allow for the
deposition of several different atomistic species.

\section*{Acknowledgments}
The authors thank Len Sander for many useful conversations. The authors
also thank Russel Caflisch and Christian Ratsch for helpful remarks;
in particular for comments that ultimately lead to the approximation
given by Eq. (\ref{eapprox}). PS was supported by NSF grants
DMS-0207402  and DMS-0244419. GR was supported by a grant from the Michigan Center
for Theoretical Physics and grant from the Italian Government (PRIN project 2003,
prot.n.2003011441\_004).

\end{document}